\documentclass[reprint,amsmath,amssymb,aps]{revtex4-2}

\usepackage{graphicx}
\usepackage{dcolumn}
\usepackage{bm}
\usepackage{here}
\graphicspath{{./figures/}}

\usepackage{color}
\usepackage[colorlinks=true,citecolor=blue,urlcolor=blue,pdfnewwindow]{hyperref}

\begin{document}

\preprint{APS/123-QED}

\title{High gradients at SRF photoinjector cavities with low RRR copper cathode plug screwed to the cavity back wall}
\thanks{Work performed in the framework of R\&D for future accelerator operation modes at the European XFEL and financed by the European XFEL GmbH.}

\author{E. Vogel} \email{elmar.vogel@desy.de}
\author{J. Sekutowicz}
\author{D. Bazyl}
\author{T. Büttner}
\author{B. van der Horst}
\author{J. Iversen}
\author{D. Klinke}
\author{A. Muhs}
\author{\\D. Reschke}
\author{S. Sägebarth}
\author{P. Schilling}
\author{M. Schmökel}
\author{L. Steder}
\author{J.-H. Thie}
\author{H. Weise}
\author{M. Wiencek}

\affiliation{Deutsches Elektronen-Synchrotron DESY, Notkestraße 85, D-22607 Hamburg, Germany}

\date{\today}

\begin{abstract}
In recent years we increased the typical maximum peak field on axis gradients obtained in L-band superconducting RF (SRF) photoinjector cavities at vertical tests to around 55\,MV/m. This was achieved with niobium cathode plugs directly screwed to the cavity back wall omitting an RF choke filter and a load lock system for cathodes. Copper demonstrated being a suitable cathode material in normal conducting injector cavities used at X-Ray Free Electron Lasers (XFELs) operating with pulsed RF. In this article we present the first experimental confirmation that peak field on axis gradients around 55\,MV/m and beyond can be achieved in L-band SRF photoinjector cavities with copper cathode plugs screwed to the cavity back wall. We view this as a major milestone for the development of a high gradient photoinjector operating continuous wave (CW).
\end{abstract}

\maketitle

\section{Introduction}
High gradient photoinjector cavities enable ``pancake'' emission of beams and direct matching into subsequent L-band SRF linacs \cite{FEL2019.wea01}. An additional buncher section as required in other setups \cite{FEL2014.THP042} can be omitted. For facilities operating CW and High-Duty-Cycle (HDC), L-band SRF photoinjector cavities provide the required gradients. They are the first choice for the planned HDC operation of the European XFEL \cite{FEL2013.TUOCNO04,FEL2014.MOP067}. 

\begin{figure}[b]
	\centering
	\includegraphics*[width=0.6\columnwidth]{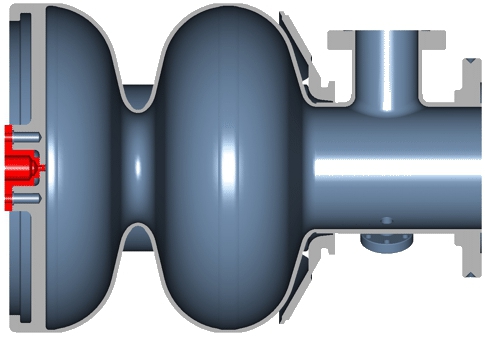}
	\caption{Photoinjector cavity with a cathode plug (red) screwed into a hole on the cavity backside.}
	\label{fig:gun_cavity_with_plug}
\end{figure}

Our cavity design (see Fig.\,\ref{fig:gun_cavity_with_plug}) consists of a cathode plug screwed to the cavity back wall \cite{IPAC2013.mopfi003,physrevstab.16.123402}. Indium is used for sealing and for good thermal contact. Well established surface treatments like high pressure rinsing (HPR) can be applied after cathode insertion in a clean room. A main prerequisite are photocathodes which are robust against the exposure to air and the usual SRF cavity cleaning procedures. Consequently our cathode material choice is mainly limited to metals.

\begin{figure}[pt]
	\centering
	\includegraphics*[width=1\columnwidth]{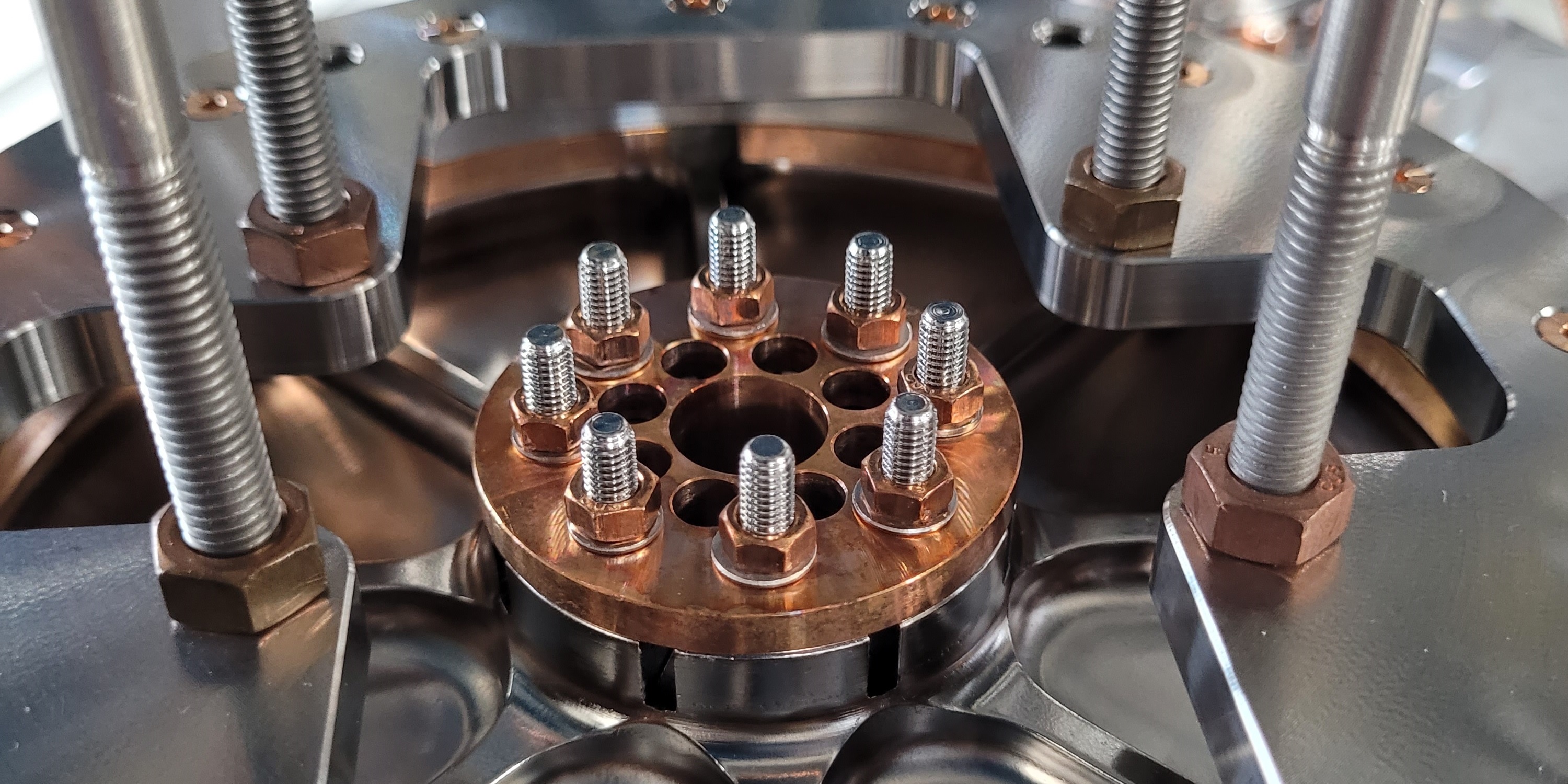}
	\caption{Copper cathode plug in the SRF cavity back wall.}
	\label{fig:copper_plug_at_cavity_back_wall}
\end{figure}

In recent years we developed the surface treatment procedures for our cavities showing now typical maximum peak field on axis gradients around 55\,MV/m in vertical tests with niobium cathode plugs \cite{IPAC2023.wepa145}.

However, the quantum efficiency (QE) of bulk niobium is around \( 10^{-7} \) and too low to use it as the emitter \cite{physrevstab.8.010701,JournalAppPhys.1.2008389}. For the European XFEL, we need a QE of \( 10^{-4} \) or higher to generate 100\,pC electron bunches with a repetition rate of at least 100\,kHz. Copper is our preferred cathode choice. It has a moderate work function of 4.6\,eV, excellent RF and thermal characteristics and the tendency causing dark current is minimal. Surface laser cleaning and additional surface treatments on copper cathodes offers the possibility of achieving a QE of at least \( 10^{-4} \) {\cite{FEL2014.THP030,PhysRevSTAB.18.033401}.

\section{The RF field in the cavity}
For beam generation and acceleration our L-band 1.6-cell SRF photoinjector cavities will be operated at 1.3\,GHz with the TM010 \mbox{$\pi$-mode}. The cathode plane is retracted from the back wall of the cavity by 450\,$\mu$m to optimize the beam dynamics near the cathode surface. This reduces the peak electric field at the cathode to 85\,$\%$ of the maximal peak electric field along the axis. To study the performance of the cavity half-cell with the cathode plug in more detail, we also measure the TM010 \mbox{0-mode} in our vertical tests. At the \mbox{$\pi$-mode} the peak electric field is almost identical in both cells. In contrast at the \mbox{0-mode}, the peak field is twice as high in the half-cell than in the full cell. Both modes differ by about 18\,MHz in frequency. The axial electric field distribution of the TM010 0- and \mbox{$\pi$-mode} is illustrated in Figure \ref{fig:field_profile} and the electric and magnetic field maps of the TM010 modes can be found in Figure \ref{fig:field_maps}.

\begin{figure}[pt]
	\centering
	\includegraphics*[width=1\columnwidth]{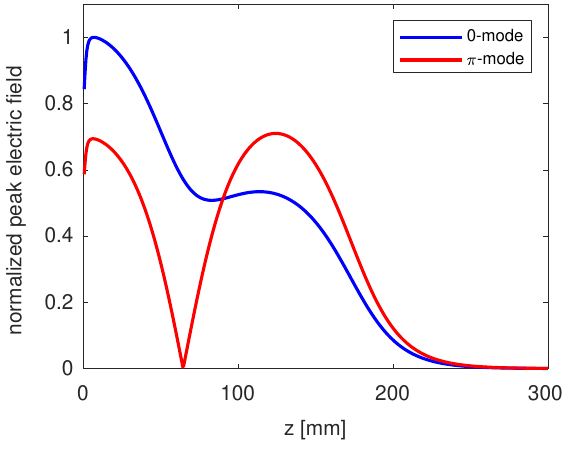}
	\caption{Distribution of the peak electric field on the cavity's axis for the TM010 0- and \mbox{$\pi$-mode}.}
	\label{fig:field_profile}
\end{figure}

\begin{figure}[pt]
	\centering
	\includegraphics*[width=1\columnwidth]{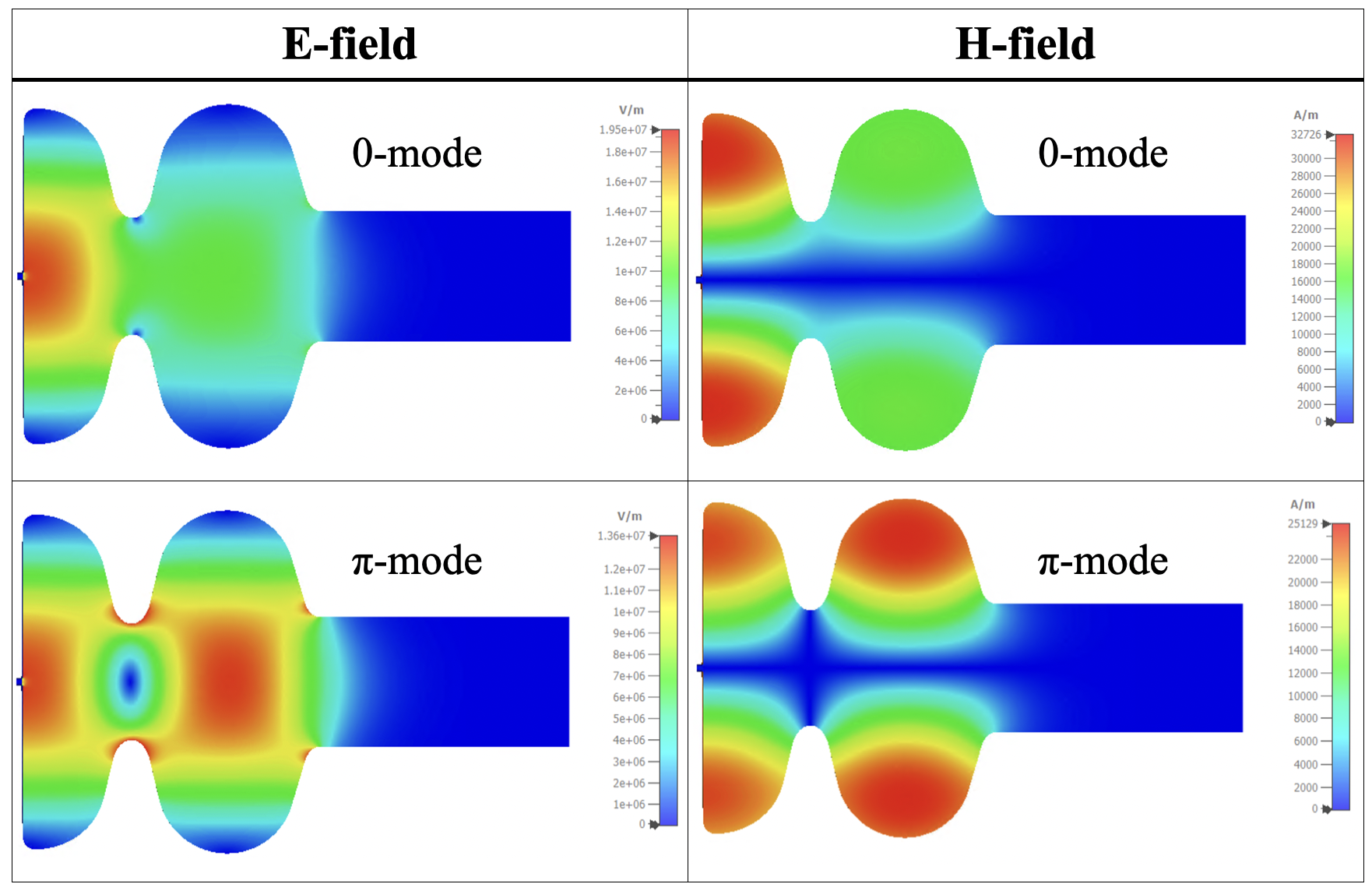}
	\caption{Illustration of the electric and the magnetic field distribution of the TM010 0- and \mbox{$\pi$-mode}.}
	\label{fig:field_maps}
\end{figure}

At the TM010 modes the surface magnetic field is zero at the radial center of the cathode, likewise the surface current. It increases with the radial offset, Fig.\,\ref{fig:cathode_H_field}. Due to the small cathode radius of 2\,mm, the surface magnetic field at the cathode is a factor of 42 lower as compared to the peak surface magnetic field in the equator region of the cavity. This combined with the good thermal conductivity of copper justified the expectation that the low dissipated RF power at the cathode should not affect the superconductivity in the remaining cavity surface.

\begin{figure}[t]
	\centering
	\includegraphics*[width=0.8\columnwidth]{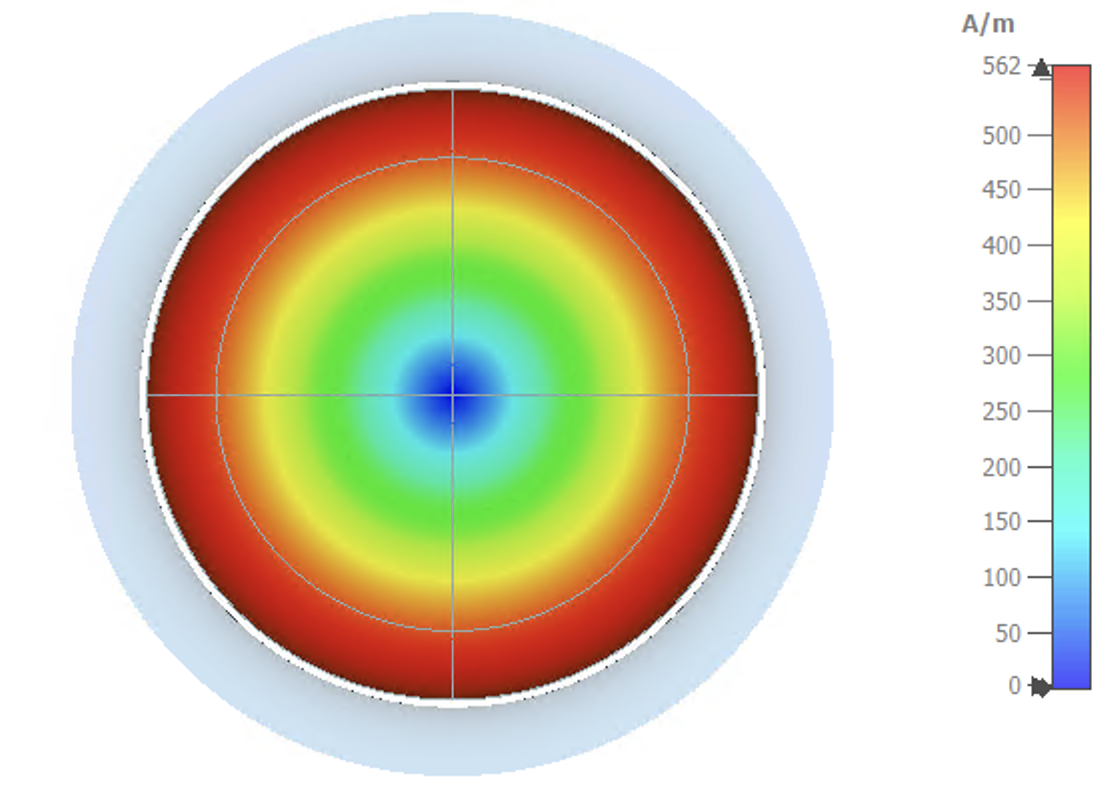}
	\caption{Magnetic field distribution at the surface of the cathode for the TM010 \mbox{$\pi$-mode}. The field amplitude is normalized for 1\,J stored energy in the photoinjector cavity.}
	\label{fig:cathode_H_field}
\end{figure}

\section{Vertical tests with copper plugs}

In May 2023 we received the most recent manufactured prototype cavities 16G09 and 16G10 after final buffered chemical polishing (BCP) treatment. Initial tests with niobium cathode plugs showed the expected maximum peak field on axis gradients around 55\,MV/m at the \mbox{$\pi$-mode} for 16G10. At the first cool down 16G09 suffered from a cold leak. Re-tightening all sealing connections and no additional cleaning was performed. Afterwards, 16G09 was leak tight but showed a significant lower gradient in the first test limited by strong radiation.

\begin{figure}[b]
	\centering
	\includegraphics*[width=0.8\columnwidth]{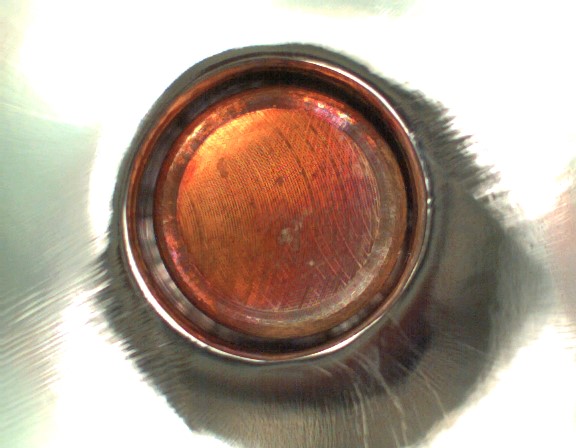}
	\caption{Photograph of the inner surface of the back wall of a photoinjector cavity with inserted copper cathode plug.}
	\label{fig:copper_plug_after_VT_insp}
\end{figure}

In August, we fitted both cavities with cathode plugs made of industrial-grade copper (Fig.\,\ref{fig:copper_plug_at_cavity_back_wall}) with comparatively low heat conductivity (residual resistivity ratio (RRR) around 30) and performed vertical tests. The cathode insertion was followed by the usual HPR and 90\,°C baking \cite{IPAC2023.wepa145}. Figure \ref{fig:copper_plug_after_VT_insp} shows the inner surface of the back wall of a SRF photoinjector cavity, photographed during a clean room optical inspection after the successful vertical tests.\\

Figure \ref{fig:G09G10CU} contains the successful vertical test results obtained with the cavities 16G09 and 16G10. Both cavities reached maximum peak field on axis gradients around 55\,MV/m in the \mbox{$\pi$-mode} and about 65\,MV/m in the \mbox{0-mode}.

We assume the difference of the maximum peak field on axis between the \mbox{$\pi$-mode} and the \mbox{0-mode} being caused by losses in the stainless steel vacuum flange connected to the cavity beam tube for the vertical tests. For a given maximum peak field on axis the remaining field is higher near the vacuum flange in the case of the \mbox{$\pi$-mode}, Fig.\,\ref{fig:field_profile}. Dedicated flanges coated with copper or niobium to verify this assumption and minimize the effect for future tests are in preparation.

\begin{figure}[pt]
	\centering
	\includegraphics*[width=1.0\columnwidth]{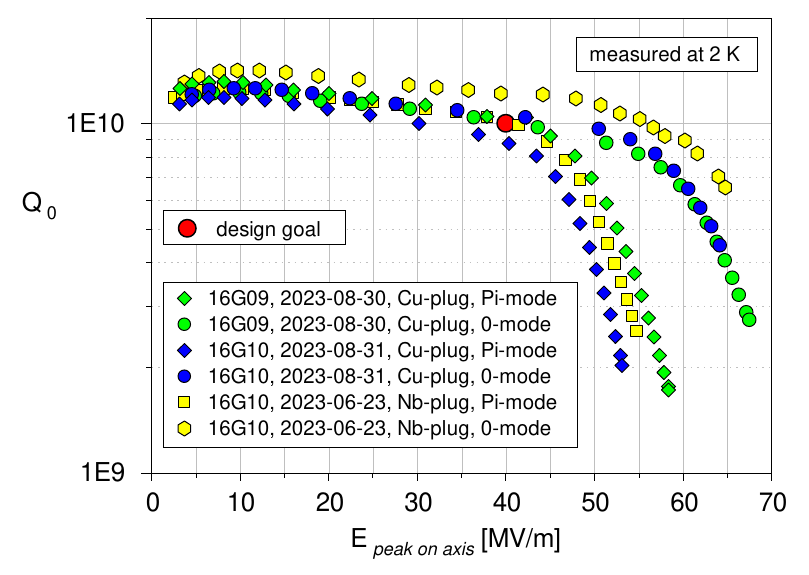}
	\caption{Q vs E curves describing the successful test results of the SRF photoinjector cavities 16G09 and 16G10 with copper cathode plugs and the first test of 16G10 with niobium cathode plug for comparison. The design goal indicated reflects the specification for the TM010 \mbox{$\pi$-mode} in terms of the Q-value and the peak electric field values.}
	\label{fig:G09G10CU}
\end{figure}

However, there is almost no performance degradation with the copper cathode plugs as compared to the niobium cathode plug. The high gradients are still achieved. The Q-values appear to be slightly lower with copper cathode plugs compared to those with niobium cathode plugs, Fig.\,\ref{fig:G09G10CU}.

We intend to try other heat treatment parameters and apply electro polishing at the cavities and using copper with high heat conductivity for the cathode plugs to reduce the Q-slope in general.

Field emission and dark current are of particular interest at photoinjector cavities. In our tests, some field emission was measurable in the first test runs. Conditioning took place and in the later tests shown in Figure \ref{fig:G09G10CU} the field emission detected decreased to irrelevant low levels. Until now, we did not implement measures to optimize the cavity cathode hole rim. We plan this for the future.\\

\section{Summary}
We proofed for the first time, that high gradients can be obtained with normal conducting cathode plugs with comparatively low heat conductivity screwed to the back wall of L-band SRF photoinjector cavities. This enables many new options for cathodes used in our cavities. Using higher-purity copper for cathode plugs enhances heat transfer from the cathode to the liquid helium bath.

\section{Acknowledgments}
The authors acknowledge the significant contributions from numerous colleagues at all institutes joining the effort for an SRF photoinjector cavity with a cathode plug directly screwed to the back wall.  Many people from industry contribute to this effort as well.

\bibliography{SRFgunCavitiesCuPlug.bib}

\end{document}